\documentstyle[epsfig]{article}

\begin{document}\large\bibliographystyle{plain}\begin{titlepage}
\hfill\begin{tabular}{l}HEPHY-PUB 703/98\\UWThPh-1998-58\\October
1998\end{tabular}\\[4cm]
\begin{center}{\Large\bf SOLVING THE SCHR\"ODINGER EQUATION FOR\\[1ex] BOUND
STATES WITH MATHEMATICA 3.0}\\
\vspace{2cm}
{\Large\bf Wolfgang LUCHA}\\[.5cm]Institut f\"ur Hochenergiephysik,\\
\"Osterreichische Akademie der Wissenschaften,\\Nikolsdorfergasse 18, A-1050
Wien, Austria\\[2cm]
{\Large\bf Franz F.~SCH\"OBERL}\\[.5cm]Institut f\"ur Theoretische Physik,\\
Universit\"at Wien,\\ Boltzmanngasse 5, A-1090 Wien, Austria\\[2.5cm]
{\bf Abstract}\end{center}
\normalsize
\noindent
Using Mathematica 3.0, the Schr\"odinger equation for bound states is solved.
The method of solution is based on a numerical integration procedure together
with convexity arguments and the nodal~theorem for wave functions. The
interaction potential has to be spherically symmetric. The solving procedure
is simply defined as some Mathematica function. The output is the energy
eigenvalue and the reduced wave function, which is provided as an
interpolated function (and can thus be used for the calculation of, e.g.,
moments by using any Mathematica built-in function) as well as plotted
automatically. The corresponding program {\tt schroedinger.nb} can be
obtained from {\tt franz.schoeberl$@$univie.ac.at}.

\vfill\noindent{\em PACS:} 03.65.Ge\normalsize\end{titlepage}
\section{Introduction}The Schr\"odinger equation is one of the fundamental
equations (of motion) in physics. Unfortunately, exact analytic solutions may
be found only in exceptional cases. A~wide area of application has been and
still is nonrelativistic potential models, which~describe bound-state
properties of hadrons, considered as bound states of quarks interacting~via
some spherically symmetric potential. Detailed discussions may be found in
Refs.~\cite{grosse80,lucha91}. Fortunately, the two-body Schr\"odinger
equation with spherically symmetric potential $V(r)$, where $r\equiv|{\bf
x}|$ and ${\bf x}$ is the relative coordinate of the constituents, can be
reduced to an ordinary differential equation for the reduced wave function
$y_{n,\ell}(r)$, describing~a bound state of radial and
orbital-angular-momentum quantum numbers $n$ and~$\ell$,~resp. In natural
units, where $\hbar=c=1$, the (nonrelativistic) two-body Schr\"odinger
equation in configuration space for the reduced wave function, which is
normalized according~to \begin{equation}\displaystyle\int\limits_0^\infty{\rm
d}r\,[y_{n,\ell}(r)]^2=1\ ,\label{eq:norm-cond}\end{equation} reads
\begin{equation}\left[\displaystyle\frac{1}{2\,\mu}
\left(-\displaystyle\frac{{\rm d}^2}{{\rm
d}r^2}+\displaystyle\frac{\ell\,(\ell+1)}{r^2}\right)+V(r)\right]
y_{n,\ell}(r)=E_{n,\ell}\,y_{n,\ell}(r)\ .\label{eq:schroedi}\end{equation}
$\ell=0,1,2,\ldots$ denotes the angular-momentum quantum number, $\mu$ the
reduced mass, $$\mu\equiv\displaystyle\frac{m_1\,m_2}{m_1+m_2}\ ,$$ and
$E_{n,\ell}$ is the energy eigenvalue, with $n=0,1,\ldots$ counting the
number of nodes of the bound-state wave function within $(0,\infty)$, which
corresponds to the radial excitations.

Here, a Mathematica \cite{mathemat97} notebook is constructed for the
numerical solution of~the reduced Schr\"odinger equation. The solution of
differential equations with Mathematica consumes more computational time than
with Fortran since the Mathematica program is not compiled. On the other
hand, it is rather complicated to handle graphics within Fortran. Moreover,
using Fortran one has to write a program for any new potential,~to compile
it, and to link it. Thus, it is much more tedious to study various
potentials~with Fortran than with Mathematica. In Mathematica, it is rather
easy to define a function, to calculate matrix elements, and to use graphics
tools.

The outline of this paper is to discuss first the method of finding the
solutions,~then to demonstrate the application of Mathematica, and, finally,
to give, in the Appendix, the program listing. (It may be obtained as
Mathematica notebook {\tt schroedinger.nb} from {\tt
franz.schoeberl@univie.ac.at}.) The method applied here is based on
Ref.~\cite{falk85}.

\section{Method of Solution}Rewriting Eq.~(\ref{eq:schroedi}) in more
convenient form, the differential equation to be integrated~is
\begin{equation}y''_{n,\ell}(r)=\left[V_{\rm
eff}(r)-\varepsilon_{n,\ell}\right]y_{n,\ell}(r)\
,\label{eq:schroed1}\end{equation} with the effective potential $$V_{\rm
eff}(r)\equiv 2\,\mu\,V(r)+\displaystyle\frac{\ell\,(\ell+1)}{r^2}$$ and the
scaled energy eigenvalue $$\varepsilon_{n,\ell}\equiv 2\,\mu\,E_{n,\ell}\ .$$

For the energy eigenvalue $E$ to be bounded from below, the potential $V(r)$
has~to~be, in any case, less singular than $-1/r^2$. Moreover, for potentials
$V(r)$ such that $r^2\,V(r)$~is analytic (which implies that $V(r)$ is less
singular than $1/r^2$), the general (normalizable) solution of the
differential equation (\ref{eq:schroed1}) is given as a power-series
expansion of~the~form $$y(r)\propto r^{\ell+1}\,[1+O(r)]\ .$$ The asymptotic
behaviour of the not normalized reduced wave function $y(r)$ for $r\to 0$ is
therefore $$\lim_{r\to 0}y(r)=r^{\ell+1}\ .$$

The normalization condition (\ref{eq:norm-cond}) forces the reduced wave
function $y(r)$ to approach zero when $r\to\infty$. Consequently, the main
idea of finding the energy eigenvalues~$E$~and corresponding reduced wave
functions $y$ is to perform a systematic scan for increasing values of
$\varepsilon$ in Eq.~(\ref{eq:schroed1}), looking for those values of
$\varepsilon$ which allow $y(r)\to 0$ for $r\to\infty$. Without loss of
generality, near the origin $y$ may be assumed to be positive: $y(0+)\ge 0$.
\begin{itemize}\item For $\varepsilon$ low enough, $V_{\rm
eff}(r)-\varepsilon$ is certainly positive and thus $y(r)\to+\infty$ for
$r\to\infty$.\item Increasing $\varepsilon$, the divergence of $y(r)$ for
$r\to\infty$ will weaken.\item Increasing $\varepsilon$, $V_{\rm
eff}(r)-\varepsilon$ will become negative in certain regions of $r$. If this
region~is large enough, it may happen that $y(r)$ vanishes for $r\to\infty$
(cf. Fig.~\ref{fig:y-veff}): the lowest bound-state energy $E_{0,\ell}$ for
given $\ell$ has been found.
\begin{figure}[h]\begin{center}\psfig{figure=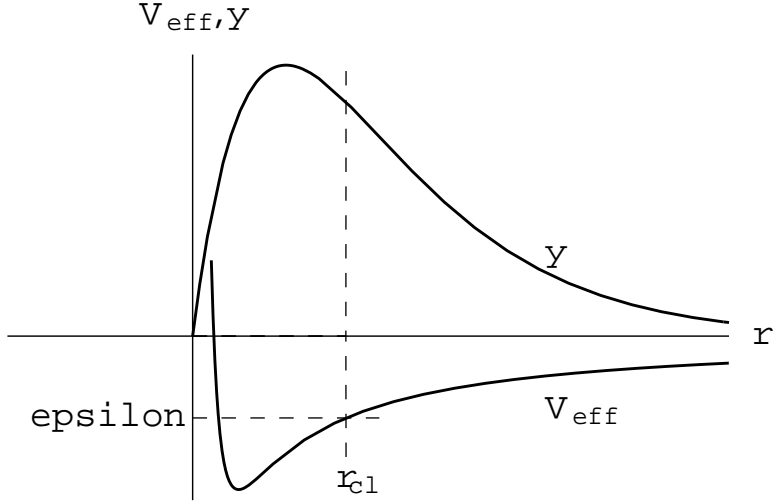}
\caption{Qualitative behaviour of potential $V_{\rm eff}(r)$ and reduced
ground-state wave function $y(r)$}\label{fig:y-veff}\end{center}\end{figure}
\item Increasing $\varepsilon$, $y(r)$ will cross $0$ somewhere and behave
like $y(r)\to-\infty$ for $r\to\infty$.\item Increasing $\varepsilon$
further, it may happen that $y(r)$ approaches 0 from below for $r\to\infty$:
the energy eigenvalue $E_{1,\ell}$ of the first radial excitation for given
$\ell$ has been found.\end{itemize}

In more detail our procedure works as follows: In principle, the integration
of Eq.~(\ref{eq:schroed1}) starts at the origin. For singular potentials,
however, it has to start at some~value~$r=\delta$ close to but different from
the origin and to respect, of course, the boundary~conditions
\begin{eqnarray}y(\delta)&=&\delta^{\ell+1}\ ,\nonumber\\[1ex]
y'(\delta)&=&(\ell+1)\,\delta^\ell\ .\label{eq:boundary}\end{eqnarray} First,
note that there is a classical turning point such that $V_{\rm
eff}(r)>\varepsilon_{n,\ell}$ for all $r>r_{\rm cl}$; the classical turning
point $r_{\rm cl}$ is the largest value of $r$ solving the equation $V_{\rm
eff}(r)=\varepsilon_{n,\ell}$. Secondly, note
that$$\mbox{sign}\;y''(r)=\mbox{sign}\;y(r)\quad\mbox{for all}\ r>r_{\rm cl}\
.$$ This means, the reduced wave function $y$ is convex (concave) if it is
positive (negative). Thus, at some point $r_>>r_{\rm cl}$, $y(r_>)>0$ and
$y'(r_>)>0$ implies $y(r)\to+\infty$ for $r\rightarrow\infty$ while
$y(r_>)<0$ and $y'(r_>)<0$ implies $y(r)\to-\infty$ for $r\rightarrow\infty$.
Clearly, for both~cases the integration can be stopped.

The level of excitation found in the course of the integration procedure is
identified, according to the well-known nodal theorem, by the number $n$ of
nodes of the (reduced) radial wave function: the ground-state wave function
has no node at all ($n=0$), while the wave function of some radial excitation
has a finite number of nodes ($n=1,2,\dots$).

In order to locate a desired bound state, define an interval for the energy
$\varepsilon$, by~fixing appropriate lower and upper bounds $E_{\rm L}$ and
$E_{\rm U}$, resp. The routine starts to integrate, with the Runge--Kutta
method \cite{abramow}, the differential equation~(\ref{eq:schroed1}) at
$r=\delta$, respecting~the boundary conditions (\ref{eq:boundary}) and using,
for the energy $\varepsilon$, the arithmetic mean $(E_{\rm L}+E_{\rm U})/2$.
It counts the number of nodes, $n$, detected within a prescribed interval
($E_{\rm L},E_{\rm U}$)~during the integration and changes $E_{\rm L}$ and
$E_{\rm U}$ appropriately in an iterative procedure: if these bounds are
chosen badly, $\varepsilon$ converges to that bound which lies next to the
true~energy; in order to cover nevertheless the desired bound state, the
corresponding bound has~to be changed.\footnote{\normalsize\ The program
listing in the Appendix gives more details.}

\section{Applications}The notebook is called {\tt schroedinger.nb}. It
identifies $r_{\rm cl}$ by determination of the~local minimum of the
effective potential at largest $r$. For numerical reasons, it uses an upper
bound on this minimum by adding one stepsize $h$ to it. Of course, for some
potentials, like all pure power-law potentials, this local minimum may be
determined analytically. For instance, for some potential of the form
$V(r)=r^k$, $k\in N$, the minimum resides~at
$$r=\left[\frac{\ell\,(\ell+1)}{k\,\mu}\right]^{1/(k+2)}\ .$$ For more
complicated potentials, the local minimum has to be determined numerically.
This is done by the module {\tt xwmil1}; the arguments of the latter are
orbital excitation~$\ell$, stepsize $h$ of the numerical integration, as well
as stepsize {\tt weit} and starting value~{\tt xrat} of the minimum search.
Both $\ell$ and $h$ are taken over from the calling procedure {\tt schroe},
which calls {\tt xwmil1} with stepsize {\tt weit=0.5} and guessed minimum at
{\tt xrat=20}.~(These values are rather good starting values for the minimum
search; they may be changed~in the module {\tt schroe}.)

Upon starting the notebook {\tt schroedinger.nb}, one is provided with
explanations~of the usage as well as a description of the definitions used.
The procedure goes as follows:\\[1.5ex]{\em Input\/}: {\tt
schroedinger.nb}.\\[1.5ex]{\em Output\/}: description of usage and
definitions.\\[1.5ex]{\em Input\/}: potential {\tt vl[r\_]:=\dots}; e.g., for
the harmonic-oscillator potential {\tt vl[r\_]:=r\^{}2}.\begin{tabbing}{\em
Input\/}: \={\tt schroe[0,20,2,1,0.01,1,1]}. The arguments of this routine
are:\\\>lower bound $E_{\rm L}$ on the energy eigenvalue $\varepsilon$ in
chosen units of energy ($=0$),\\\>upper bound $E_{\rm U}$ on the energy
eigenvalue $\varepsilon$ in chosen units of energy ($=20$),\\\>number $n$ of
nodes = radial excitations ($=2$),\\\>angular momentum $\ell$ = orbital
excitations ($=1$),\\\>stepsize $h$ in chosen units of energy
($=0.01$),\\\>masses $m_1$, $m_2$ in chosen units of energy
($m_1=m_2=1$).\end{tabbing}{\em Output\/}: energy eigenvalue $E_{n,\ell}$ in
chosen units of energy, etc.\\[1.5ex]The above example generates as output
the seeked numerical results and---if desired---a plot of the (not
normalized) reduced wave function $y$ (here called {\tt yschr}) as function
of $r$ (here called {\tt x}):\\[1.5ex]{\tt E = 12.99999714, L = 1, N =
2,\\Integration steps = 589, h = 0.01, del = 0.001, el = 0, eu = 20,\\Largest
x, upper integration limit, XMAX = 5.881,\\Smallest x, lower integration
limit, XMIN = 0.001.\\The not normalized reduced wave function is {\tt
yschr[x]}.\\The normalization factor is given
by:\\1/NIntegrate[yschr[x]\^{}2,\{x,del,xmax\}]}

\begin{figure}[h]\begin{center}\psfig{figure=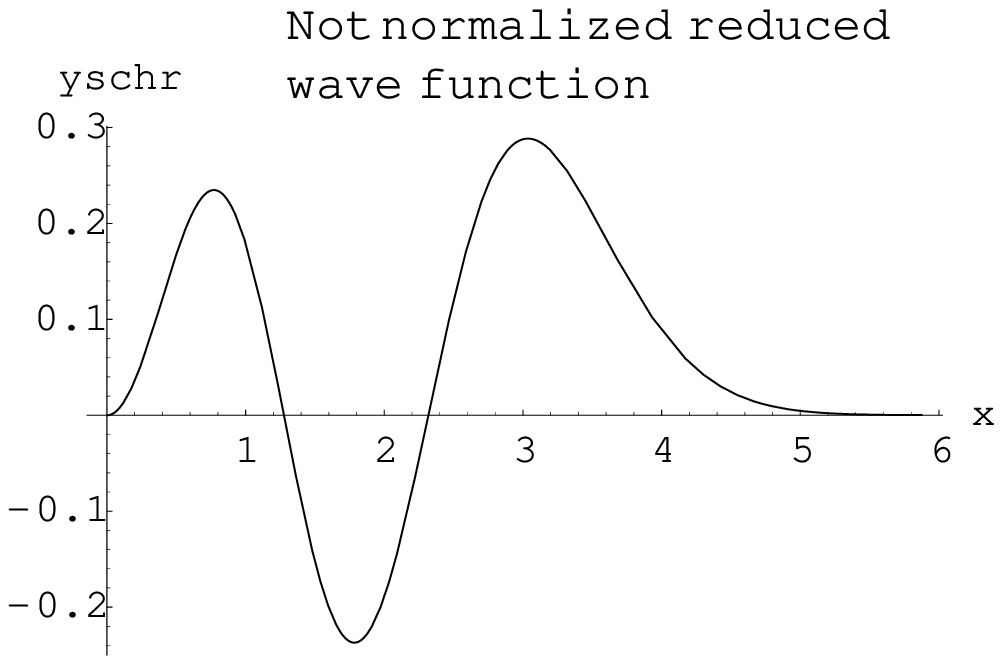}\\[0.5cm]
\end{center}\end{figure}

We are now able to use Mathematica built-in functions to evaluate matrix
elements like $\langle 1/r\rangle$.\begin{tabbing}{\em Input\/}: \={\tt
NIntegrate[1/r yschr[r]\^{}2,\{r,del,xmax\}]/}\\\>{\tt
NIntegrate[yschr[r]\^{}2,\{r,del,xmax\}]}.\end{tabbing}{\em Output\/}: {\tt
0.625982}.\\

Table~\ref{tab:osci} lists results of a test run for the harmonic-oscillator
potential while Table~\ref{tab:h-atom} shows results for the Coulomb
potential for varying stepsizes $h$ and starting points~{\tt del}.

Obviously, $y(0)$ has to vanish for all states while $y'(0)$ is nonvanishing
only for $\ell=0$ states---which provides a (trivial) additional consistency
check. Of course, the~required computational time depends strongly on the
desired accuracy.

In summary, the Mathematica notebook developed here provides an
easy-to-handle procedure for computing energies and wave functions of
(nonrelativistic) bound states.

\newpage\begin{table}[h]\caption{Energy eigenvalues, moments, as well as wave
function and its derivative at the origin for~the harmonic-oscillator
potential $V(r)=r^2$. The values chosen for the parameters are
$m_1=m_2=1\;\mbox{GeV}$, $h=0.01\;\mbox{GeV}^{-1}$, $E_{\rm
L}=0\;\mbox{GeV}$, and $E_{\rm U}=20\;\mbox{GeV}$. The exact energies are
$E_{n,\ell}=4\,n+2\,\ell+3\;\mbox{GeV}$;~the discrepancies of the computed
values only reflect the $h$-dependent accuracy of our numerical
method.}\label{tab:osci}
\begin{center}\begin{tabular}{ccrcccc}\hline\hline\\[-1.5ex]$n$&$\ell$&
$E_{n,\ell}$[GeV]&$\langle 1/r\rangle$ [GeV]&$\langle 1/r^2\rangle$ [GeV$^2$]&
$[y(0)]^2$ [GeV]&$[y'(0)]^2$ [GeV$^3$]\\[1ex]\hline\\[-1.5ex]
0&0&2.999997&1.1284&1.9977&0.0000&2.2567\\
1&0&7.000003&0.9403&1.9966&0.0000&3.3851\\
2&0&10.999999&0.8369&1.9958&0.0000&4.2314\\[1ex]
0&1&4.999995&0.7523&0.6667&0.0000&0.0000\\
1&1&9.000001&0.6770&0.6667&0.0000&0.0000\\
2&1&12.999997&0.6260&0.6667&0.0000&0.0000\\[1ex]
0&2&7.000003&0.6018&0.4000&0.0000&0.0000\\
1&2&10.999999&0.5588&0.4000&0.0000&0.0000\\
2&2&15.000005&0.5266&0.4000&0.0000&0.0000\\[1ex]
\hline\hline\end{tabular}\end{center}\end{table}

\vspace*{-.2ex}

\begin{table}[h]\caption{Energy eigenvalues, moments, and derivative of the
wave function at the origin for the~1S,~2S, and 2P states of the Coulomb
potential $V(r)=-1/r$ as function of the stepsize $h$. The starting~point
$\delta$ is calculated from $\delta=h/10$; the numerical values chosen for
the parameters are $m_1=m_2=1\;\mbox{GeV}$, $E_{\rm L}=-1\;\mbox{GeV}$, and
$E_{\rm U}=+1\;\mbox{GeV}$. The corresponding exact results are given in
parentheses.}\label{tab:h-atom}
\begin{center}\begin{tabular}{cllllll}\hline\hline\\[-1.5ex]
State&1S&&2S&&2P&\\[1ex]\hline\\[-1.5ex]$h$
[GeV$^{-1}$]&0.1&0.05&0.1&0.05&0.1&0.05\\[1ex]\hline\\[-1.5ex]$-E_{n,\ell}$
[GeV]&0.249973&0.249996&0.062496&0.062496&0.062504&0.062504\\
&(0.25)&&(0.0625)&&(0.0625)&\\[1ex] $\langle 1/r\rangle$
[GeV]&0.4999&0.4999&0.1250&0.1250&0.1250&0.1250\\
&(0.5)&&(0.125)&&(0.125)&\\[1ex] $\langle 1/r^2\rangle$
[GeV$^2$]&0.4947&0.4974&0.06185&0.06221&0.02082&0.02082\\
&(0.5)&&(0.0625)&&(0.02083)&\\[1ex] $[y'(0)]^2$
[GeV$^3$]&0.4897&0.4949&0.06123&0.06190&2.$10^{-6}$&4.$10^{-7}$\\
&(0.5)&&(0.0625)&&(0)&\\[1ex]\hline\hline\end{tabular}\end{center}\end{table}

\newpage
\appendix
\section{Program Listing}
\begin{tabbing}
\=($*$\\
\=Author: \=Franz F. Sch\"oberl\\
\>\>Institute for Theoretical Physics\\
\>\>University of Vienna\\
\>\>Boltzmanng. 5\\
\>\>A-1090 Vienna\\
\>\>E-MAIL: Franz.Schoeberl$@$Univie.ac.at
\end{tabbing}
\begin{center}
{\Large SOLVING THE SCHR\"ODINGER EQUATION FOR BOUND STATES WITH MATHEMATICA
3.0}
\end{center}
This software is based on:\\
''Solving the Schr\"odinger Equation for Bound States" by\\
P. Falkensteiner, H. Grosse, Franz F. Sch\"oberl, and P. Hertel\\
Computer Physics Communication 34 (1985) 287-293.\\
\noindent
The energy and the reduced radial wave function for a bound state with given
number of nodes n0 and angular momentum l is calculated. The potential has to
be spherically symmetric.\\

\begin{tabbing} 
 \=Usage:\\
 \> el ..... \=lower bound of the energy,\\
 \> eu ..... \>upper bound of the energy,\\
 \> n0 ..... \>radial excitations, number of nodes,\\
 \> l ...... \>orbital excitations,\\
 \>	feh .... \>error on the energy, built in as feh=0.00001,\\
 \>h ......\>integration stepsize, determines the number of the integration
steps\\ \>	 \> and thus the accuracy of the determined energy\\ \>	m1,
m2 ..... constituent masses,\\ \> schroe[el,eu,n0,l,h,m1,m2] ..... calling
procedure
\end{tabbing}
 The output reduced wave function (not normalized) is called yschr[x].\\
 The plot of the wave function is called yschrplot.\\
 Example, harmonic oscillator:\\
 (1) Define the potential: vl[x\_]:= x\^\,2\\
 (2) Start solving the Schr\"odinger equation:
schroe[-1,20,1,2,0.1,0.336,0.336]\\ The above procedure will solve the
equation for the first radial and second orbital excitation.\\ The equation
to be solved is (with $\hbar = c = 1,\quad \mu = \frac{m_1m_2}{m_1 +
m_2}$):\\ $[\frac{1}{2\mu}(-\frac{d^2}{dr^2}+\frac{\ell (\ell
+1)}{r^2})+V(r)]y_{n,\ell}(r)$, $\mbox{yschr[x]}$ is the not normalized
$y_{n,\ell}(r)$,\\ $\psi_{n,\ell ,m}(r)=\frac{y_{n,\ell}(r)}{r}Y_{\ell
,m}(\Omega)$, $\displaystyle\int_0^\infty (y_{n,\ell}(r))^2dr = 1.$\\

\noindent
The accuracy can be increased by decreasing h, this increases the number of
integration steps. The higher the number of integration steps the more
accurate the eigenvalues as well as the eigen functions. The reduced wave
function will be plotted in addition. A measure for the accuracy is also the
shape of the wave function. It should vanish for the largest numerical x,
otherwise one has to decrease h.\\ If you run schroe[el,eu,n0,l,h,m1,m2] you
automatically
will be asked if you like to plot the reduced wave function.\\$*$)
\begin{tabbing} 
\=12345 \= \kill
\>($*$\\
\> xwmil1 calculates the minimum of the potential most to the right. The
minimum\\
\> is called xwmin. xrat is some guessed x-value most to the right\\
\> (here I have used x-rat = 20). wei1 is the stepsize of the minimum
search\\
\> (here I have used wei1 = 0.5).\\
\>$*$)\\
\>\>xwmil1[l1\_, h1\_, wei1\_, xrat1\_, ww1\_] :=\\
\>\> Module[\{l = l1, h = h1, weit = wei1, xrat = xrat1, ww = ww1\},\\
\>\> del = h/10;\\
\>\> xs = xrat;\\
\>\> If[xs $<$ del, Goto[11]];\\
\>\> If[l $<$ 1, Goto[3]];\\
\>\> Label[1];\\
\>\> If[xs$<$weit + del, xs = del; Goto[2]];\\
\>\> ms = ww*vl[xs] + l$*$(l + 1)/xs\^\,2;\\
\>\> rs = ww*vl[xs + weit] + l$*$(l + 1)/(xs + weit)\^\,2;\\
\>\> ls = ww*vl[xs - weit] + l$*$(l + 1)/(xs - weit)\^\,2;\\
\>\> If[rs$>$ms \&\& ms$>$ls, xs = xs - weit; Goto[1]];\\
\>\> If[rs$<$ms \&\& ls$>$ms, xs = xs + weit; Goto[1]];\\
\>\> If[rs = ls , Goto[2]];\\
\>\> If[rs$>$ms \&\& ls$>$ms, If[weit$<$h, Goto[2]];\\
\>\> weit = weit/10; Goto[1]];\\
\>\> Print["SOMETHING IS WRONG, ANALYZE POTENTIAL"];\\
\>\> Goto[10];\\
\>\> Label[3];\\
\>\> Label[1 a];\\
\>\> If[xs$<$weit + del, xs = del; Goto[2]];\\
\>\> ms = ww*vl[xs];\\
\>\> rs = ww*vl[xs + weit];\\
\>\> ls = ww*vl[xs - weit];\\
\>\> If[rs$>$ms \&\& ms$>$ls, xs = xs - weit; Goto[1 a]];\\
\>\> If[rs$<$ms \&\& ls$>$ms, xs = xs + weit; Goto[1 a]];\\
\>\> If[rs = ls , Goto[2]];\\
\>\> If[rs$>$ms \&\& ls$>$ms, If[weit$<$h, Goto[2]];\\
\>\> weit = weit/10; Goto[1 a]];\\
\>\> Print["SOMETHING IS WRONG, ANALYZE POTENTIAL"];\\
\>\> Goto[10];\\
\>\> Label[11];\\
\>\> Print["XMIN TOO SMALL"];\\
\>\> Goto[10];\\
\>\> Label[2];\\
\>\> xwmin = xs;\\
\>\> Label[10];\\
\>\> Return["end"]],\\
\>\\ 
\>($*$\\
\> {The program schroe}\\
\>$*$)\\ 
\>\>schroe[el1\_, eu1\_, n01\_, l1\_, h1\_, m1\_, m2\_]:=\\
\>\>Module[\{el = el1$*$ww, eu = eu1$*$ww, n0 = n01, l = l1, h = h1, w1 =
m1,\\
\>\>w2 = m2\},\\
\>($*$\\
\>\>Determining the minimum of the potential most to the right.\\
\> $*$)\\
\>\>ww = 2$*$w1$*$w2/(w1 + w2);\\
\>\>xwmil1[l, h, 0.5, 20, ww];\\
\>\>xwmin = xwmin + h;\\
\>\>del = h/10;\\
\>\>feh = 0.00001$*$ww;\\
\> ($*$\\
\>\>Defining diffl = Veff - e, $\mbox{y}^{\prime\prime}$ = (Veff - e) y. 
\\
\>$*$)\\
\>\>diffl[xxx\_, l\_, een\_] := ww$*$vl[xxx] + l$*$(l + 1)/xxx\^\,2 - een;\\
\>\>Label[300];\\
\>\>seh = eu - el;\\
\>\>eps = (el + eu)/2;\\
\>($*$\\
\>\>Starting the integration with the boundaries y(del) = del\^\,(l + 1),\\
\>\>$\mbox{y}'$(del) = (l + 1)del\^\,l, and with the energy eps equal to the
arithmetic\\
\>\>mean of the preceding lower and upper, bounds of energy el and eu.\\
\> $*$)\\ 
\>\>x = del;\\
\>\>y = x\^\,(l + 1);\\
\>\>yp =(l + 1)$*$x\^\,l;\\
\>\>yold = 1;\\
\>\>n0x = 0;\\
\>($*$\\
\>\>If the desired accuracy (prescribed error feh) has been obtained, the
bound\\
\>		 \>state energy is taken as the arithmetic mean of the last
el and eu.\\
\> $*$)\\
\>\>If[seh$<$feh, Goto[1]];\\
\>($*$\\
\>\>Integrating $\mbox{y}^{\prime\prime}$ = (Veff - e)y one step h further
with the Runge - Kutta\\
\>\> method\\
\> $*$)\\
\>\>Label[2];\\
\>\>a1 = yp$*$h; b1 = diffl[x, l, eps]$*$h$*$y;\\
\>\>a2 = (yp + b1/2)$*$h; hh = diffl[x + h/2, l, eps]$*$h;\\
\>\>b2 = hh$*$(y + a1/2); a3 = (yp + b2/2)$*$h; b3 = hh$*$(y + a2/2);\\
\>\> a4 = (yp + b3)$*$h;\\
\>\>x = x + h; u2 = diffl[x, l, eps]; b4 = u2$*$h$*$(y + a3);\\
\>\>y = y + (a1 + 2$*$a2 + 2$*$a3 + a4)/6;\\
\>\>yp = yp + (b1 + 2$*$b2 + 2$*$b3 + b4)/6;\\
\>($*$\\
\>\>Counting the number of nodes by n0x until the prescribed n0 is
reached.\\
\> $*$)\\
\>\> If[y$*$yold$>$0, Goto[3]];\\
\>\> n0x = n0x + 1;\\
\>\> If[n0x$>$n0, Goto[4]];\\
\>\> Label[3];\\ 
\>\> yold = y;\\
\>($*$\\
\>\> If the following condition is not fullfilled, x is greater then the
classical\\
\>\> turning point (u2 has the value of Veff - eps at the point x).\\
\> $*$)\\
\>\> If[(u2$<$0 $||$ x$<$xwmin), Goto[2]];\\
\>($*$\\
\>\>If (after stating that x is greater than the classical turning point)
y$*$yp is\\
\>\>greater than 0 (i.e., y and yp have the same sign), one is sure that y
goes to\\
\>\> infinity, without having additional nodes. Otherwise one has to
integrate\\
\>\> further,\\
\> $*$)\\
\>\> z = y$*$yp;\\
\>\> If[z$<$0, Goto[2]];\\
\>($*$\\
\>\> If y goes to infinity, a new el is established by eps.\\
\> $*$)\\
\>\> el = eps;\\
\>\> Goto[300];\\
\>($*$\\
\>\> If nox exceeds n0, a new eu is established by eps.\\
\> $*$)\\
\>\> Label[4];\\
\>\> eu = eps;\\
\>\> Goto[300];\\
\>($*$\\
\>\> In the following lines the wave function y is calculated using the
above\\
\>\> calculated bound state energy (the last eps) by the same method as
above. In\\
\>\> addition y is stored in feld1 at x which is stored in xcoord, and the
number\\ \>\> of integration steps is counted by j.\\
\> $*$)\\ 
\>\> Label[1];\\
\>\> ep = eps;\\
\>\> j = 0;\\ 
\>\> Label[20];\\
\>\> j = j + 1;\\
\>\> feld1[j] = y;\\
\>\> xcoord[j] = del + (j - 1)$*$h;\\
\>\> j1 = j;\\
\>\> xmax = xcoord[j1];\\
\>($*$\\
\>\> Integrating $\mbox{y}^{\prime\prime}$ = (Veff - e)y one step h further
with the\\
\>\> Runge - Kutta method\\
\> $*$)\\ 
\>\> a1 = yp$*$h; b1 = diffl[x, l, eps]$*$h$*$y;\\
\>\> a2 =(yp + b1/2)$*$h; hh = diffl[x + h/2, l, eps]$*$h;\\
\>\> b2 = hh$*$(y + a1/2); a3 = (yp + b2/2)$*$h; b3 = hh$*$(y + a2/2);\\
\>\> a4 =(yp + b3)$*$h;\\
\>\> x = x + h; u2 = diffl[x, l, eps]; b4 = u2$*$h$*$(y + a3);\\
\>\> y = y + (a1 + 2$*$a2 + 2$*$a3 + a4)/6;\\
\>\> yp = y + (b1 + 2$*$b2 + 2$*$b3 + b4)/6;\\
\>\> If[y$*$yold$>$0, Goto[30]];\\
\>\> n0x = n0x + 1;\\
\>\> If[n0x$>$n0, Goto[40]];\\
\>\> Label[30];\\
\>\> yold = y;\\
\>\> If[(u2$<$0 $||$ x$<$xwmin), Goto[20]];\\
\>\> z = y$*$yp;\\
\>\> If[z$<$0, Goto[20]];\\
\>\> Label[40];\\
\>($*$\\
\>\> The reduced radial wave function yschr obtained from the
interpolation\\
\>\> of the data is stored in feld1 and in xcoord\\
\> $*$)\\ 
\>\> yschr = Interpolation[Table[\{xcoord[j],feld1[j]\}, \{j, 1, j1\}]];\\
\>\> xmax = xcoord[j1];\\
\> ($*$\\
\>\> Output of the resulting energy eigenvalue and the input data.\\
\> $*$)\\
\>\> Print[\\
\>\> StyleForm["E = ",\\
\>\> FontColor -$>$ RGBColor[0.996109, 0, 0], FontWeight -$>$ "Bold",\\
\>\> FontSize -$>$ 16],\\ 
\>\> StyleForm[N[ep/ww, 10],\\
\>\> FontColor -$>$ RGBColor[0.996109, 0, 0], FontWeight -$>$ "Bold",\\
\>\> FontSize -$>$ 16], ",",\\
\>\> StyleForm["L = ",\\
\>\> FontColor -$>$ RGBColor[0.996109, 0, 0], FontWeight -$>$ "Bold",\\
\>\> FontSize -$>$ 16],\\
\>\> StyleForm[l,\\
\>\> FontColor -$>$ RGBColor[0.996109, 0, 0], FontWeight -$>$ "Bold",\\
\>\> FontSize -$>$ 16], ",",\\
\>\> StyleForm["N = ",\\
\>\> FontColor -$>$ RGBColor[0.996109, 0, 0], FontWeight -$>$ "Bold",\\
\>\> FontSize -$>$ 16],\\
\>\> StyleForm[n0,\\
\>\> FontColor -$>$ RGBColor[0.996109, 0, 0], FontWeight -$>$ "Bold",\\
\>\> FontSize -$>$ 16], ",",\\
\>\> Integrationsteps = ", j1, ",", "h = ", h, ",", " del =",\\ 
\>\> del, ",", " el = ", el1, ",", " eu = ", eu1, ",",\\
\>\> StyleForm["Largest x, upper integration limit, XMAX = ",\\
\>\> FontColor -$>$ RGBColor[0, 0, 0.996109], FontWeight -$>$ "Bold",\\
\>\> FontSize -$>$ 16],\\
\>\> StyleForm[xmax,\\
\>\> FontColor -$>$ RGBColor[0, 0, 0.996109], FontWeight -$>$ "Bold",\\
\>\> FontSize -$>$ 16], ",",\\
\>\> StyleForm["Smallest x, lower integration limit, XMIN = del = ",\\
\>\> FontColor -$>$ RGBColor[0, 0, 0.996109], FontWeight -$>$ "Bold",\\
\>\> FontSize -$>$ 16],\\
\>\> StyleForm[del,\\
\>\> FontColor -$>$ RGBColor[0, 0, 0.996109], FontWeight -$>$ "Bold",\\
\>\> FontSize -$>$ 16],".",\\
\>\> StyleForm["The reduced not normalized wave function is yschr[x].\\
\>	\>	 The normalizationfactor is given by:\\
\>\>		 1/NIntegrate[yschr[x]\^\,2,\{x,del,xmax\}]",\\
\>\> FontColor -$>$ RGBColor[0.996109, 0.500008, 0.250004],\\
\>\> FontWeight -$>$ "Bold", FontSize -$>$ 16]];\\
\>($*$\\ 
\>\> Preparing the plot of the reduced wave function yschr[x]\\
\> $*$)\\ 
\>\> zz = InputString["You like to plot the (not normalized) reduced wave\\
\>\> function? Type yes and click OK, otherwise click just OK "];\\
\>\> If[zz == "yes", yschrplot = Plot[yschr[x], \{x, del, xmax\},\\
\>\> PlotStyle -$>$ GrayLevel[0], AxesLabel -$>$ \{"x", "yschr"\},\\
\>\> DefaultFont -$>$ \{"Times-Bold", 12\},\\
\>\> Background -$>$ RGBColor[0.996109, 0.996109, 0],\\
\>\> PlotLabel -$>$ FontForm["Not normalized\\
\>\> reduced wave function", \{"Helvetica-Bold", 14\}]],\\
\>\> Print[StyleForm["OK NO PLOT", FontWeight -$>$ "Bold", FontSize -$>$
18,\\
\>\> FontColor -$>$ RGBColor[0, 0.500008, 0]]]];\\
\>\> Return[]]
\end{tabbing}\end{document}